\documentstyle[twocolumn, amssymb, prl, aps, epsfig]{revtex}

\begin{document}
\title{Coherent Backscattering of Light by Cold Atoms}
\author{G. Labeyrie$^{\ast }$, F. de Tomasi$^{\dagger }$, J.-C. Bernard$^{\ast }$,
C. A. M\"{u}ller$^{\ast }$, C. Miniatura$^{\ast }$\ and R. Kaiser$^{\ast }$}
\address{* Institut Non Lin\'{e}aire de Nice, UMR 6618, 1361 route des Lucioles,
F-06560\ Valbonne.\\
$\dagger $\ now at Dipartimento di Fisica, Universita di Lecce, via Arnesano.}
\date{\today}
\maketitle

\begin{abstract}
Light propagating in an optically thick sample experiences multiple
scattering. It is now known that interferences alter this propagation,
leading to an enhanced backscattering, a manifestation of weak localization
of light in such diffuse samples. This phenomenon has been extensively
studied with classical scatterers. In this letter we report the first
experimental evidence for coherent backscattering of light in a laser-cooled
gas of Rubidium atoms.
\end{abstract}


Transport of waves in strongly scattering disordered media has received much
attention during the past years when it was realized that interference can
dramatically alter the normal diffusion process\cite
{sharvin,Sheng,mesoscopic}. In a sample of randomly distributed scatterers,
the initial direction of the wave is fully randomized by scattering and a
diffusion picture seems an appropriate description of propagation when the
sample thickness is larger than the scattering mean free path\cite
{Chandrasekhar}. This model neglects all interference phenomena and predicts
a transmission of the medium inversely proportional to sample thickness.
This is the familiar Ohm's law. However, interferences may have dramatic
consequences such as a vanishing diffusion constant\cite{anderson58}. In
this situation, the medium behaves like an insulator (strong localization).
Such a disorder induced transition has been reported for microwaves and for
light\cite{wiersma97}. In fact, even far from this insulating regime,
interferences already hamper the diffusion process (weak localization). This
has been demonstrated in coherent backscattering (CBS) experiments. Upon
coherent illumination of a static sample, a random speckle pattern is
generated. This pattern is washed out by configuration averaging except in a
small angular range around the backscattering direction where constructive
interferences originating from reciprocal light paths enhance diffuse
reflection from the sample\cite{tiggelen90}. This effect has been observed
for light in a variety of different media such as suspensions of powder
samples, biological tissues or Saturn's rings \cite{CBSdivers}, as well as
for acoustic waves\cite{tourin97}. Among other interesting features such as
universal conductance fluctuations\cite{scheffold98} or lasing in random
media\cite{cao99}, CBS is a hallmark of coherent multiple scattering.

Atoms as scatterers of light offer new perspectives. The achievements of
laser cooling techniques\cite{lasermanipulation,Cornelldavis95}\ in the last
decade now allow to manipulate and control samples of quantum scatterers.
Cold atoms are unique candidates to move the field of coherent multiple
scattering to a fully quantum regime (quantum internal structure,
wave-particle duality, quantum statistical aspects). For instance, the
coupling to vacuum fluctuations (spontaneous emission) is responsible for
some unusual properties of the scattered light (elastic and inelastic
spectra \cite{mollow69,gao94}). Also, information encoding in atomic
internal states can erase interference fringes like in some ''which-path''
experiments \cite{itano98}. Furthermore, it is now possible to implement
situations where the wave nature of the atomic motion is essential \cite
{lasermanipulation,Cornelldavis95}.

In our experiment, the scattering medium is a laser-cooled gas of Rubidium
atoms which constitutes a perfect monodisperse sample of strongly resonant
scatterers of light. The quality factor of the transition used in our
experiment is $Q=\nu _{at}/\Delta \nu _{at}\approx 10^{8}$ (D2 line at $%
\lambda =c/\nu _{at}=780\,$nm, intrinsic resonance width $\Delta \nu
_{at}=\Gamma /2\pi =6\,$MHz). The scattering cross section can thus be
changed by orders of magnitude by a slight detuning of the laser frequency $%
\nu _{L}$. This is a new situation compared to the usual coherent multiple
scattering experiments where resonant effects, if any, are washed out by the
sample polydispersity. Moreover in our sample the duration $\tau _{D}$
(delay time) of a single scattering event largely dominates over the free
propagation time between two successive scattering events : for on-resonant
excitation ($\delta =\nu _{L}-\nu _{at}=0$), this delay is of the order of $%
\tau _{D}\approx 2/\Gamma =50$ ns corresponding to free propagation of light%
{\em \ }over $15$ m in vacuum. In such a situation, particular care must be
taken to observe a CBS effect. Indeed, when atoms move, additional
phaseshifts are involved. Configuration averaging will only preserve
constructive interference between reciprocal waves if the motion-induced
optical path change $\Delta x$ does not exceed one wavelength \cite
{golubentsev84}. A rough estimate is $\Delta x=v_{rms}\tau _{D}<\lambda $, a
criterium which can be written in the more appealing form $kv_{rms}<\Gamma $%
. Thus, for resonant excitation, the Doppler shift must be small compared to
the width of the resonance. For Rubidium atoms illuminated by resonant
light, one finds $v_{rms}<4.6\,$m/s corresponding to a temperature $T=200\,$%
mK.\ Much lower temperatures are easily achieved by laser cooling thus
allowing observation of interference features in multiple scattering.
However, up to now, only incoherent effects in multiple scattering, like
radiation trapping \cite{holstein47}, have been investigated in cold atomic
vapors\cite{fioretti98}.

We prepare our atomic sample by loading a magneto-optical trap (MOT) from a
dilute vapor of Rubidium 85 atoms\cite{lasermanipulation} (magnetic gradient 
$\nabla B\approx 7\,$G/cm, loading time{\em \ }$t_{load}\approx 0.7\sec $).
Six independent trapping beams are obtained by splitting an initial laser
beam slightly detuned to the red of the trapping transition (power per beam $%
30\,$mW, FWHM diameter $2.8\,$cm, Rubidium saturation intensity $%
I_{sat}\approx 1.6\,$mW/cm$^{2}$, $\delta \approx -3\Gamma $). The repumper
is obtained by two counterpropagating beams from a free running diode laser
tuned to the $F=3\rightarrow F^{\prime }=3$ transition of the D2 line.
Fluorescence measurements yield $N\approx 10^{9}$ atoms corresponding to a
spatial density $n_{at}\approx 2\times 10^{9}$ cm$^{-3}$ at the center of
the cloud (gaussian profile, FWHM diameter $\approx 7\,$mm). The velocity
distribution of the atoms in the $\,$trap has been measured by a
time-of-flight technique to be $v_{rms}\approx 10\,$cm/s, well below the
limit imposed by the above velocity criterium. To observe coherent
backscattering (CBS) of light, we alternate a CBS measurement phase with a
MOT preparation phase. During the CBS phase, the magnetic gradient and
trapping beams of the MOT are switched off\ (residual power per beam $%
0.2\,\mu $W). The CBS probe beam (FWHM $\approx 6\,$mm, spectral width $%
\Delta \nu _{L}\approx 1$ MHz) is resonant with the closed trapping
transition of the D2 line : $F=3\rightarrow F^{\prime }=4$. A weak probe is
used to avoid saturation effects (power $80\,\mu $W, on-resonant saturation $%
s=0.1$).\ The optical thickness of the sample, measured by transmission, is $%
\eta \approx 4$ and remains constant, within a few percent, during the whole
duration of the CBS measurement phase ($2.5\,$ms). The corresponding
extinction mean free path $\ell \approx 2\,$mm is consistent\ with an
estimation deduced from our fluorescence measurements, taking a scattering
cross-section at resonance{\em \ }$\sigma _{res}=\,3\lambda ^{2}/2\pi $.

\begin{figure}
\centerline{\epsfig{file=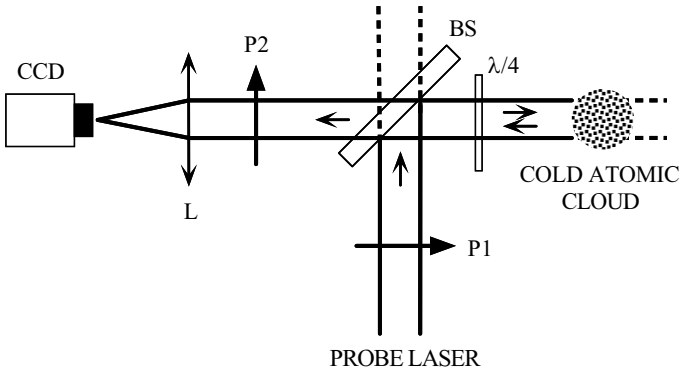,width=0.5\textwidth}}
\caption{The CBS detection scheme. P1, P2 : polarizers; $\lambda /4$ :
quarter-wave plate; BS : beam-splitter (T $=90\%$); L : analysis lens ($%
f=500 $ mm).}
\label{fig1}
\end{figure}

The CBS detection setup is shown in Fig.\ref{fig1}. It involves a cooled CCD camera
in the focal plane of a converging lens ($f=500\,$mm). A polarization
sensitive detection scheme, generally allowing to eliminate the single
scattering contribution\cite{albada85}, is used for signal recording in
various polarization channels. For a linear incident polarization, we record
the scattered light with (linear) polarization parallel (''parallel''
channel) or orthogonal (''orthogonal'' channel) to the incident one. We also
use a circular incident polarization by inserting a quarter-wave plate
between the beam-splitter and the sample. In the ''helicity preserving''
channel the detected polarization is circular with the same helicity (sign
of rotation of the electric field referenced to the wave propagation
direction) as the incident one : as an example, no light is detected in this
channel in the case of\ the back-reflection by a mirror. This is the channel
mostly used in previous studies, because it allows to eliminate the single
scattering contribution (for dipole-type scatterers). The ''helicity
non-preserving'' channel is obtained for a detected circular polarization
orthogonal to the previous one. Teflon or dilute milk samples were used to
find the exact backward direction, to cross-check the polarization channels
and to test the angular resolution of our set-up. During the MOT phase
(duration $10\,$ms), probe beam and detection scheme are switched off while
the MOT is switched on again to recapture the atoms. After this phase a new
atomic sample is reproduced. The whole sequence is repeated for a typical
duration of $1$ min with{\em \ }a detected flux typically about $1800$
photons/pixel/sec. A ''background'' image, representing less than 10\% of
the full signal level (due mainly to scattering from the repumper by hot
atoms in the cell), is substracted from the ''CBS'' image to suppress stray
light contributions.

\begin{figure}
\caption{}
\label{fig2}
\end{figure}

Fig.~\ref{fig2} (color image in appendice) shows the CBS images obtained from our laser-cooled Rubidium vapor in
the various polarization channels. We clearly observe enhanced
backscattering in all four polarization channels whereas for a thick teflon
sample we only found pronounced cones in the polarization preserving
channels. This enforces the idea that low scattering\ orders are dominant in
our experiment\cite{Albada87} which is not surprising considering the
relatively small optical thickness of our sample. The intensity enhancement
factors, defined as the ratio between the averaged intensity scattered in
exactly backward direction and the large angle background are 1.11,
1.06,1.08 and 1.09 for the helicity preserving, helicity non-preserving,
orthogonal and parallel channels respectively. The detected light
intensities in these channels, normalized to that of the linear parallel
channel, are 0.76, 0.77, 0.54 and 1. A closer look at Fig.\ref{fig2}d 
reveals that
the cone exhibits a marked anisotropy in the (linear) parallel polarization
channel: the cone is found to broader in the (angular) direction parallel to
the incident polarization. This effect has already been observed in
classical scattering samples and is also a signature of low scattering
orders \cite{Albada87}{\em .}

\begin{figure}
\centerline{\epsfig{file=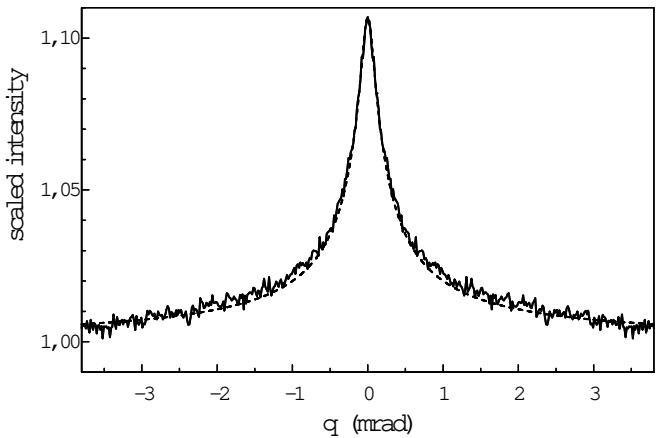,width=0.55\textwidth}}
\caption{Atomic CBS cone in the helicity non-preserving channel. The
experimental profile (solid line) is a section of the 2D image of Fig. 2a
(after angular averaging). The dashed curve is a fit by a model assuming
only double scattering, with the mean free path determined experimentally
and the enhancement factor the only adjustable parameter.}
\label{fig3}
\end{figure}

For a more quantitative analysis of the CBS cone, we report in Fig.\ref{fig3} 
a section of image \ref{fig2}a (helicity non-preserving channel), taken after an
angular average was performed on the data (this procedure is justified when
the cone is isotropic, as in Fig.\ref{fig2}a). The measured cone width\ $\Delta
\theta $ is about $0.57\,$mrad, nearly six times larger than our
experimental resolution of $0.1\,$mrad. Taking into account the experimental
resolution, we compared our data to a calculation (dotted line) involving
only double scattering \cite{tiggelen90}. The experimental value $\ell
\approx 2\,$mm for the mean free path was used in the calculation, leaving
the enhancement factor as free parameter. Even though the assumptions
underlying this theoretical model (isotropic double scattering,
semi-infinite medium) are rather crude in our case, the shape of the CBS
cone is nicely reproduced. We plan, in further studies, to investigate in
more details the contributions of different scattering orders by carefully
analyzing the CBS cone shape.

One important aspect in CBS\ studies has always been the enhancement factor
in the backscattering direction, due to the constructive interferences
between reciprocal paths. In the helicity preserving channel, this
enhancement factor is known to be 2 for independent scattering by classical
scatterers\cite{CBSdivers}, as single scattering can be ruled out in that
polarization channel. In our experiment with cold atoms however,\ we measure
a backscattered enhancement of 1.06, clearly less than 2 ! This reduction
cannot be attributed to the experimental resolution, as we have measured
enhancement factors on milk (using a dilution giving the same cone width as
the atomic one) of 1.8.\ However, in our situation several processes could
reduce the cone contrast. The first one is single scattering, which does not
contribute to CBS. Due to the presence of several Zeeman sublevels in the
groundstate of Rubidium atoms, Raman processes, i.e. light scattering with
change of the atomic internal sublevel, have to be considered. In such
events, the polarization of the scattered light differs from the incident
polarization and single scattering is not eliminated even in the helicity
preserving channel. Another consequence of the atom's internal structure is
a possible imbalance between the amplitudes of the reciprocal waves : atoms
in different internal states can have different scattering cross sections
(resulting from different Clebsch-Gordan coefficients). They can thus be
seen as partial polarizers which can imbalance the amplitude of the paths
which interfere for CBS. Furthermore finite-size effects should also be
taken into account. Indeed our sample does not have the standard slab
geometry and the gaussian shape of the probe beam is known to reduce the
enhancement factor. We are currently investigating these effects to
determine their respective magnitudes for our situation. Also, some more
subtle phenomena might play an additional role in the cone reduction. For
instance, with classical scatterers, the radiated and the incident light
have identical frequencies (elastic scattering). This is no longer true for
atoms for which the resonant fluorescence spectrum displays inelastic
structures in addition to the usual elastic component\cite{mollow69,gao94}.
Because of Raman scattering, even in the weak saturation limit (weak probe
intensity), atoms have a non-negligible probability to undergo inelastic
scattering \cite{gao94}. The role of these rather complex spectral
properties in coherent backscattering has yet to be studied both
theoretically and experimentally.

In summary we have reported the first observation of coherent backscattering
of light by a sample of laser cooled atoms. These first results indicate
that in our system low scattering orders are dominant, as expected from
optical thickness measurements. The exact value of the enhancement factor
and the precise shape of the cone is not yet fully understood and requires
more experimental and theoretical investigations. Further experiments will
include studies of the effect of the probe beam intensity (which determines
the amount of inelastic scattering) and detuning. Detuning the laser
frequency from the atomic resonance leads to an increased mean free path $%
\ell =1/n_{at}\sigma$ . Indeed, we already observed that the
measured width $\Delta \theta $ of the coherent backscattering cone
decreases when the probe frequency is detuned from resonance, as expected
from the scaling $\Delta \theta \varpropto \lambda /\ell $\cite{tiggelen90}.
It would be very interesting to extend these experiments to new regimes.
Weak and strong localization of light in gaseous Bose-Einstein condensates
and of atomic matter waves in random optical potentials certainly present a
great challenge for the near future.

We would like to thank the CNRS and to the\ PACA Region for financial
support. We also thank the POAN Research Group. Finally, we would like to
deeply thank D. Delande, B. van Tiggelen and D.Wiersma for many stimulating
discussions.


\begin{references}
\bibitem{sharvin}  Sharvin, D.Yu., \& Sharvin,Yu.V., JETP Lett. {\bf 34},
272-275 (1981).

\bibitem{Sheng}  Scattering and Localization of Classical Waves in Random
Media, P. Sheng, Eds. (World Scientific, Singapore, 1990).

\bibitem{mesoscopic}  Mesoscopic Quantum Physics, E. Akkermans, G.
Montambaux, J.-L. Pichard \& J. Zinn-Justin, Eds., Elsevier Science B.V.
(North Holland, Amsterdam, 1995).

\bibitem{Chandrasekhar}  Chandrasekhar, S., Radiative transfer (Dover, New
York, 1960).

\bibitem{anderson58}  Anderson, P.W., Phys.\ Rev. {\bf 109}, 1492-1505
(1958).

\bibitem{wiersma97}  Wiersma, D.S., Bartolini, P., Lagendijk, A. \& Righini,
R. Nature {\bf 390}, 671-673 (1997); Gresillon, S., et al., Phys. Rev. Lett. 
{\bf 82}, 4520-4523 (1999);{\em \ }Genack, A.Z. \& Garcia, N., Phys.\ Rev.\
Lett. {\bf 66}, 2064-2067 (1991).

\bibitem{tiggelen90}  van Tiggelen, B.A., Lagendijk, A. \& Tip, A., J.
Phys.: Condens. Matter {\bf 2}, 7653-7677 (1990).

\bibitem{CBSdivers}  Wiersma, D.S., van Albada, M.P., van Tiggelen, B.A. \&
Lagendijk, A., Phy. Rev. Lett. {\bf 74}, 4193-4196 (1995); Yoo, K.M., Tang,
G.C. \& Alfano, R.R., Appl. Opt. {\bf 29}, 3237-3239 (1990); Mishchenko,
M.I., Astrophys. J. {\bf 411}, 351-361 (1993).

\bibitem{tourin97}  Tourin, A., Derode, A., Roux, P., van Tigelen, B.A. \&
Fink, M., Phys. Rev. Lett. {\bf 79}, 3637-3639 (1997).

\bibitem{scheffold98}  Scheffold, F. \& Maret, G. Phys.\ Rev.\ Lett. {\bf 81}%
, 5800-5803 (1998).

\bibitem{cao99}  Cao, H., et al., Phys.\ Rev.\ Lett. {\bf 82}, 2278-2281
(1999).

\bibitem{lasermanipulation}  Laser Manipulation of Atoms and Ions, E.
Arimondo, W.D.\ Phillips \& F.\ Strumia, Eds. (North Holland, Amsterdam,
1992).

\bibitem{Cornelldavis95}  Anderson, M.H., Ensher, J.R., Matthews, M.R.,
Wieman, C.E. \& Cornell, E.A., Science {\bf 269}, 198-201 (1995). Davis,
K.B., et al.,\ Phys. Rev. Lett. {\bf 75}, 3969-3973 (1995).

\bibitem{mollow69}  Mollow, B.R., Phys.\ Rev. {\bf 188}, 1969-1975 (1969).

\bibitem{gao94}  Gao, B., Phys.\ Rev.\ A {\bf 50}, 4139-4156 (1994).

\bibitem{itano98}  Itano, W.M., et al., Phys.\ Rev.\ A {\bf 57}, 4176-4187
(1998).

\bibitem{golubentsev84}  Golubentsev, A.A., Sov.\ Phys.\ JETP {\bf 59},
26-34 (1984).

\bibitem{holstein47}  Holstein, T., Phys.\ Rev. {\bf 72}, 1212-1233 (1947).

\bibitem{fioretti98}  Fioretti, A., Molisch, A.F., Muller, J.H., Verkerk, P.
\& Allegrini, M., Optics Comm. {\bf 149}, 415-422 (1998).

\bibitem{albada85}  van Albada, M.P. \& Lagendijk, A.,\ Phys.\ Rev. Lett. 
{\bf 55}, 2692-2695 (1985); Wolf, P.-E. \& Maret, G., Phys.\ Rev. Lett. {\bf %
55}, 2696-2699 \ (1985).

\bibitem{Albada87}  van Albada, M.P., van der Mark, M. \& Lagendijk, A.,\
Phys.\ Rev. Lett. {\bf 58}, 361-364 (1987).


\end{references}
\end{document}